\documentclass[
  10pt,
  sort&compress,
  5p
]{elsarticle}

\usepackage{amsmath,amsfonts}
\usepackage{amsthm}

\usepackage{graphicx}
\usepackage{booktabs}
\usepackage{threeparttable}
\usepackage{siunitx}
\usepackage[font=small,labelfont=bf]{caption}
\usepackage{balance}
\usepackage{subcaption}
\usepackage[inline,shortlabels]{enumitem}
\setlist{nosep}

\usepackage{titlesec}
\titlespacing*{\section}{0pt}{1.2ex}{0.8ex}
\titlespacing*{\subsection}{0pt}{1.0ex}{0.6ex}

\usepackage[dvipsnames]{xcolor}
\PassOptionsToPackage{
  colorlinks=true,
  linkcolor=MidnightBlue,
  citecolor=MidnightBlue,
  urlcolor=MidnightBlue
}{hyperref}
\usepackage{hyperref}

\newcommand{\nt}[4]{\int \limits_{#1}^{#2}~\mathrm{d}#3~{#4}} 
\newcommand{\dd}{\mathrm{d}} 						          

\makeatletter
\def\ps@pprintTitle{%
  \let\@oddhead\@empty
  \let\@evenhead\@empty
  \let\@oddfoot\@empty
  \let\@evenfoot\@oddfoot}
\makeatother

\journal{~}

\begin{document}

\hypersetup{ 
  colorlinks=true,
  linkcolor=MidnightBlue,
  citecolor=MidnightBlue,
  urlcolor=MidnightBlue
}

\begin{frontmatter}

\title{HawkesRank: \\Event-Driven Centrality for Real-Time Importance Ranking}

\author[1]{Didier Sornette\corref{cor}}
\author[1,2]{Yishan Luo\corref{cor}}
\author[1]{Sandro Claudio Lera\corref{cor}}

\cortext[cor]{Corresponding authors: \texttt{\{luoys2020,leras\}@sustech.edu.cn} and \texttt{dsornette@ethz.ch}}
    
\address[1]{Institute of Risk Analysis, Prediction and Management, Southern University of Science and Technology, Shenzhen, China}
\address[2]{Warwick Business School, University of Warwick, Coventry, UK}

\begin{abstract}
Quantifying influence in networks is important across science, economics, and public health, yet widely used centrality measures remain limited:
they rely on static representations, heuristic network constructions, and purely endogenous notions of importance, while offering little semantic connection to observable activity.
We introduce HawkesRank, a dynamic framework grounded in multivariate Hawkes point processes that models exogenous drivers (intrinsic contributions) and endogenous amplification (self- and cross-excitation). 
This yields a principled, empirically calibrated, and adaptive importance measure. 
Classical indices such as Katz centrality and PageRank emerge as mean-field limits of the framework, clarifying both their validity and their limitations.
Unlike static averages, HawkesRank measures importance through instantaneous event intensities, enabling prediction, transparent endo–exo decomposition, and adaptability to shocks.
Using both simulations and empirical analysis of emotion dynamics in online communication platforms, we show that HawkesRank closely tracks system activity and consistently outperforms static centrality metrics.
\end{abstract}


\end{frontmatter}

\section*{Introduction}

We live in an era of information overload, where attention has become an increasingly scarce resource \cite{simon1996designing}.  
To navigate this complexity, we rely on ranking systems to assess relevance and prioritize options.  
As a result, rankings have become ubiquitous in modern society, playing a central role in decision-making across domains.  
University and academic rankings shape the distribution of talent, 
rankings of stocks and financial instruments guide the allocation of capital, 
and website search rankings have been shown to influence election outcomes by swaying undecided voters \cite{epstein2015}.  
Rankings therefore have a profound impact on how resources are distributed, raising fundamental questions about how relevance should be judged and value determined.

Many ranking systems are built upon network representations of interactions. 
Within these networks, centrality measures provide systematic tools for identifying the most important or influential nodes.  
Different metrics capture different structural aspects of connectivity.  
Among them, eigenvector-based measures, such as Katz centrality and PageRank, form a particularly influential class, 
emphasizing that relevance accrues through association with already influential nodes.
Katz centrality captures this by summing over all walks in the network with an attenuation factor that discounts longer paths,  
while PageRank operationalizes a similar idea through a random-walk model that ranks web pages \citep{brin1998}.

By translating connectivity into interpretable scores of influence, eigenvector-based centralities have found applications across domains 
to rank
countries by economic fitness \citep{hidalgo2009},  
financial institutions by systemic risk \citep{eisdorfer2022,battiston2012},  
students by popularity \citep{Lera2025},  
politicians by influence \citep{miller2015,dubois2014},  
individuals in crime networks by criminal activity \citep{bright2015,morselli2013},  
startups by expected success \citep{bonaventura2020,xu2020},
law firms by win rate \cite{Mojon2025},
scientific papers and researchers by impact \citep{diallo2016,senanayake2015,yan2011},  
and athletes by performance in sports analytics \citep{radicchi2011,castellano2019}.  
They have also been used 
to identify central neurons in brain networks \citep{fletcher2018,lohmann2010},  
to analyze protein interaction networks \citep{ivan2011},  
and to assess species importance in ecological systems \citep{allesina2009}.

Despite their success, classical centrality measures suffer from at least five interdependent limitations.  
First, they inadequately capture exogenous (henceforth \textit{exo}) drivers of relevance.  
Eigenvector-based centralities rely on a self-referential logic, assigning importance to nodes connected to other important nodes.
Such endogenous (henceforth \textit{endo}) formulations neglect activity generated independently of the network.  
Although Katz centrality includes an exo baseline term, it is typically assumed homogeneous or omitted altogether,  
limiting the ability to account for real-world drivers such as media exposure, advertising, policy shocks, or intrinsic appeal \citep{bonacich1987power, borgatti2005centrality, mariani2020}.  

Second, standard centrality measures are static. 
They assume fixed network structures and immutable relevance scores, making them unresponsive to changing environments or new information \citep{liao2017, lu2016}.  
In practice, rankings must adapt to shifting attention, shocks, and emergent trends, something current metrics can only achieve through repeated retraining.  
Even PageRank, though widely viewed as dynamic, is usually applied to static graphs and exhibits built-in temporal bias:  
earlier nodes are favored over newer entrants, entrenching outdated rankings and amplifying systemic feedback \citep{mariani2020}.  

Third, the network structures on which centrality measures operate are often constructed in an ad hoc manner.  
In many applications, the adjacency matrix is not directly observed but inferred from data through heuristic choices, such as temporal aggregation windows, thresholding rules, or normalization schemes.  
For example, interaction networks are frequently built by linking two events whenever they occur within a chosen time interval, or by projecting bipartite data onto one-mode networks.  
These procedures introduce multiple degrees of freedom that can substantially alter the resulting network and therefore the inferred rankings.  
As a result, centrality scores often depend as much on modeling assumptions as on the underlying data, limiting their statistical interpretability and robustness.

Fourth, classical centrality measures often lack semantic clarity: it is not well-defined what quantity is being ranked.  
Eigenvector-based scores are dimensionless and lack direct physical interpretation,  
which limits their comparability across contexts and their connection to observable outcomes.  
This weakens both the interpretability and empirical grounding of traditional rankings.  

Fifth, existing methods conflate qualitatively different types of exo influence.  
External signals often blend slowly evolving intrinsic value with transient visibility surges from advertising, social amplification, or opportunistic exposure.  
When the former dominates, exogenous activity reflects fundamental importance; when the latter prevails, rankings may be distorted by ephemeral or strategic effects.  
This distinction is central to disciplines such as  
ethics \citep{zimmerman2001intrinsic, anderson1993value},  
information systems \citep{floridi2011philosophy},  
finance \citep{abarbanell1997fundamental},  
and development economics \citep{sen1999development}, 
yet remains unresolved: intrinsic value is often abstract, context-dependent, and unobservable.  
In modern socio-economic systems, this ambiguity is increasingly exploited, as appearance-driven signals crowd out substance, and strategic signaling replaces value-based assessment.

Taken together, these limitations show that ranking in complex systems requires a framework capable of jointly addressing several intertwined challenges:
disentangling endogenous amplification from heterogeneous exogenous drivers, accommodating temporal evolution and shocks, reducing dependence on arbitrary network constructions, and distinguishing intrinsic importance from transient or strategically generated visibility.
To address these intertwined limitations, we introduce \textit{HawkesRank},  a dynamic ranking framework grounded in self-exciting multivariate Hawkes processes (SEMHP) \citep{Hawkesmulti71,EmbrechtsLini11,Hansenriv15,saichev2011,saichev2013} that explicitly disentangles endogenous amplification from heterogeneous exogenous drivers while accounting for the temporal evolution of influence.
Previous applications of SEMHP have focused on modeling bursty dynamics for trend detection and popularity prediction,  
such as the viral spread of tweets \citep{zhao2015seismic, Jiang2026},
online emotions \cite{Luo2025}, 
or the inference of hidden influence networks from social media interactions \citep{zhou2013learning}.  
Our contribution differs in that we repurpose this class of models as a general analytical tool for ranking,  
providing a principled way to separate endogenous reinforcement from exogenous contributions.  
We show that, in the static limit, HawkesRank recovers Katz centrality. 
But unlike static centrality measures, HawkesRank explicitly decomposes relevance into time-varying endo and exo components,  
where exo can include both intrinsic qualities and external shocks. 
This decomposition yields interpretable, event-based rankings that reflect real-time importance while remaining responsive to memory effects, network reinforcement, and environmental change.

\section*{Results}

\begin{figure*}[htb]
    \centering
    \includegraphics[width=0.85\linewidth]{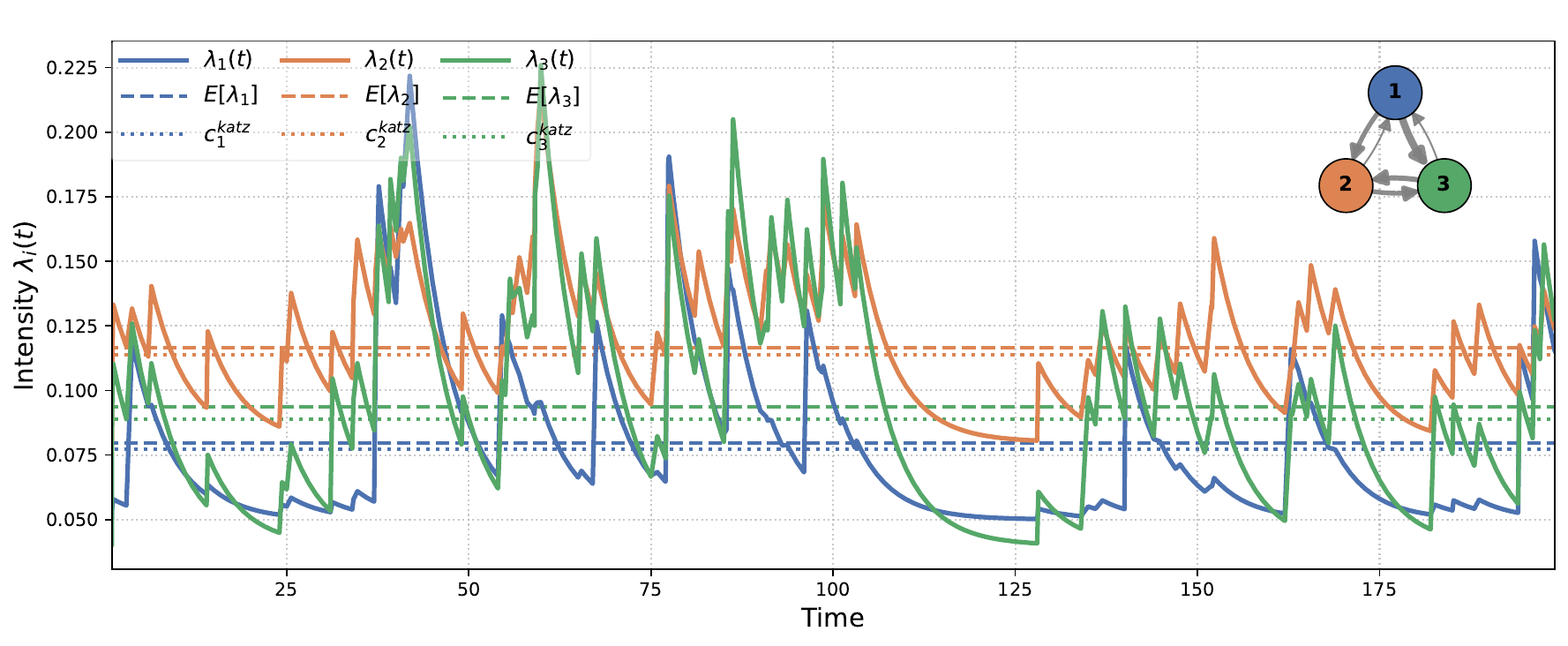}
    \caption{
    Temporal evolution of the event intensities $\lambda_i(t)$ for $i = 1,2,3$ in a three-dimensional Hawkes process.  
    The branching ratio matrix $N$ is illustrated in the top-right corner as a weighted network, where edge thickness indicates the strength of excitation between nodes.  
    The intensities evolve dynamically in response to exogenous inputs with baseline values $(\mu_1, \mu_2, \mu_3) = (0.05,\,0.08,\,0.04)$, generating endogenous bursts of activity through self- and cross-excitation.  
    As a result, the relative ranking of the intensities changes over time, with the ordering of $\lambda_1$, $\lambda_2$, and $\lambda_3$ fluctuating throughout the simulation.  
    By contrast, the first-moment Hawkes solution from Eq.~\eqref{eq:first_moment_general0} and the Katz centralities from Eq.~\eqref{eq:katz1} remain constant over time.
    }
    \label{fig:hawkes_timeseries}
\end{figure*}

\subsection*{HawkesRank: A Multivariate Hawkes Process}
\label{sec:hawkes_ranking}

We now introduce HawkesRank, which is based on the mathematics of self-excited multivariate Hawkes point processes (SEMHP).
SEMHP capture the idea that events can arise both from an exogenous background rate of activity and from cascades triggered by earlier events. 

Formally, we consider $M$ distinct \textit{event types}. 
In the network representation used by centrality metrics, these $M$ objects correspond to the $M$ nodes of the network.
For each event type $i = 1, \ldots, M$, the SEMHP specifies an intensity function $\lambda_i(t)$ that describes the instantaneous rate at which events of type $i$ are expected to occur at time $t$, that is, the probability of observing an event of type $i$ in a short interval $[t, t+\dd t)$ is equal to $\lambda_i(t) \dd t$.
In the remainder of this article, we assume this intensity is given by 
\begin{equation}
\lambda_i(t)
= 
\underbrace{
        \mu_i(t) 
        \vphantom{\sum_{j =1}^M  \sum_{k:t_k^j<t} n_{j,i} \phi( t-t_k^j )}
        }_{\text{exo}}
+ 
\underbrace{
        \sum_{j =1}^M  \sum_{k:t_k^j<t} n_{j,i} \phi( t-t_k^j )
        }_{\text{endo}}
\label{eq:Hawkes}
\end{equation}
with more general versions of the SEMHP discussed in the Methods section below.

This expression consists of two components.
The first term, the background, exogenous (exo) component $\mu_i(t)$, captures the idea that events may occur spontaneously due to external or intrinsic drivers, independently of past activity.  
For example, in the context of website visits, $\mu_i(t)$ reflects any factor that draws user attention to page $i$ from outside the network, rather than through links or referrals from other nodes.  

The second term, the endogenous (endo) component, captures the fact that past events increase the likelihood of future ones.  
Specifically, an event of type $j$ at time $t_k^j < t$ contributes to the intensity of type $i$ at time $t$ through two factors.  
First, the average fertility coefficient $n_{j,i}$ denotes the expected number of type-$i$ events directly triggered by a single event of type $j$, quantifying the average causal influence from $j$ to $i$.
As we will see below, the branching ratio matrix, here defined through its transpose $N^T = \{ n_{i,j} \}_{i,j=1}^M$, generalizes the concept of an adjacency matrix.
Second, this influence decays over time according to a normalized kernel $\phi(t) = \tfrac{1}{\tau} e^{-t/\tau}\,\mathbf{1}_{\{t > 0\}}$, where $\tau > 0$ is a memory scale.
The kernel is normalized according to $\int_0^\infty \phi(u)\,\mathrm{d}u = 1$.  
While exponential decay is a common choice, other shapes such as power laws can be used to model systems with long memory.
The scalar \textit{branching ratio} $n$, defined as the spectral radius of the matrix $N$, must satisfy $0 \leq n < 1$ to ensure that the Hawkes process is well-defined and non-explosive.

Because the intensity $\lambda_i(t)$ represents the instantaneous expected rate of events of type $i$, it provides a natural operational measure of real-time activity, and hence of importance.
A node with a higher $\lambda_i(t)$ is more likely to generate the next event and should therefore be ranked as more influential at that moment.  
For webpage ranking, $\lambda_i(t)$ represents the instantaneous rate at which page $i$ is visited, providing a direct and interpretable measure of its current relevance. In social media applications, $\lambda_i(t)$ may represent the expected posting or resharing rate of an influencer, identifying who is currently driving engagement.  
In finance, $\lambda_i(t)$ corresponds to the instantaneous trading intensity of an asset, ranking instruments by their momentary contribution to market activity.  
In neuroscience, it can be interpreted as the firing rate of a neuron, highlighting those most actively shaping network-level dynamics.  
HawkesRank formalizes this interpretation by treating the SEMHP intensities $\lambda_i(t)$ themselves as dynamic rankings, yielding a time-resolved notion of influence that evolves with the system.  
Because these intensities can be estimated directly from empirical event data, HawkesRank provides an operational benchmark against which the performance of static centrality measures can be systematically evaluated.

\subsection*{Katz Centrality: A Special Case of HawkesRank} 

First, as a useful consistency check, we show that Katz centrality and related classical measures arise as static mean-field limits of HawkesRank.
Katz centrality is motivated by the idea that a node's importance is determined recursively by the importance of its neighbors, augmented by an exogenous baseline contribution.  

Formally, let $A \in \mathbb{R}^{M \times M}$ be a nonnegative adjacency matrix, where $A_{ij} \ge 0$ quantifies the strength of the directed influence from node $j$ to node $i$.   
Denote by $\lambda_{\max}$ the spectral radius of $A$, and let $\alpha$ satisfy $0 < \alpha < 1/\lambda_{\max}$ to ensure convergence.  
Given an exogenous weight vector $\beta \in \mathbb{R}^M$, the Katz centrality vector $c^{\mathrm{katz}}$ is defined implicitly by
\begin{equation}
	c^{\mathrm{katz}} = \alpha A^\top c^{\mathrm{katz}} + \beta,
	\label{eq:katz0}
\end{equation}
with the closed-form solution
\begin{equation}
    c^{\mathrm{katz}} = (I - \alpha A^\top)^{-1} \beta.
    \label{eq:katz1}
\end{equation} 
The vector $\beta$ thus represents an exogenous source of importance, while the recursive term captures endogenous accumulation of influence through the network structure.
Choosing $\beta=\vec{1}$ recovers the classical definition where all nodes contribute equally. 
In the limiting case $\beta\to 0$, $\alpha \to 1/\lambda_{\max}$, Katz centrality converges to eigenvector centrality. 
Finally, PageRank is a normalized variant of Katz centrality, obtained by applying the Katz formulation to the row-normalized version of $A$ (see Methods for details). 

We now calculate the first moment of the SEMHP with respect to all possible stochastic realizations of the process.
Taking expectations on both sides of \eqref{eq:Hawkes}, we obtain 
\begin{subequations}
\begin{equation}
        \mathbb{E}[\lambda_i](t) 
        = 
        \mathbb{E}[\mu_i(t)] + \sum_{j =1}^M \int_{-\infty}^{t} \mathrm{d}s ~ \mathbb{E}[\lambda_j](s) n_{j,i} \phi( t-s )
        \label{eq:first_moment_general0}
\end{equation}
where $\mathbb{E}[\lambda_i](t)$ refers to the ensemble average of the stochastic intensities $\lambda_i(t)$ across all realizations of the process (Methods).
When the memory parameter $\tau$ approaches 0, the function $\phi(t)$ approaches a Dirac delta function, such that
\begin{equation}
        \mathbb{E}[\lambda_i](t) 
        =  
        \mathbb{E}[\mu_i(t)]  + \sum_{j =1}^M \mathbb{E}[\lambda_j](t)n_{j,i}.
        \label{eq:first_moment_Hawkes}
\end{equation}
\end{subequations}
This equation recovers Katz centrality in Equation~\eqref{eq:katz0} under the following conditions:
(i) the exogenous weight vector is time independent and non-stochastic, $ \mathbb{E}[\mu_i(t)] = \mu_i(t) \equiv \beta_i$;
(ii) the branching ratio matrix $N = \{ n_{j,i} \}_{j,i=1}^M$, where $n_{i,j}$ denotes the expected number of type-$i$ events directly triggered by a single type-$j$ event is equal to the adjacency matrix $A$.

For static $\mu_i$, Katz centrality is also recovered as the long-term average of $\lambda_i(t)$ for arbitrary finite values of $\tau$ (Methods).
HawkesRank thus generalizes Katz by allowing heterogeneous and explicitly time-dependent exogenous contributions $\beta_i(t)$.  
As we show below, it also provides a principled way to recover the interaction matrix $N$ directly from event data,  
in contrast to traditional approaches where the adjacency matrix $A$ is constructed heuristically and depends on arbitrary choices,  
such as the time scale over which interactions are aggregated.  
The resulting intensities $\lambda_i(t)$ define dynamic rankings that evolve through the interplay of exogenous input $\mu_i(t)$ and heterogeneous endogenous interactions encoded in $N$.  
This gives rise to punctuated bursts of activity, in contrast to the constant expectation $\mathbb{E}[\lambda_i]$ and the static baseline implied by Katz centrality (Figure~\ref{fig:hawkes_timeseries}).
HawkesRank thereby unifies classical centralities within a dynamic, event-driven framework that is more expressive, interpretable, and responsive to real-time system behavior.

\subsection*{The Importance of Dynamic Rankings}
\label{sec:dynamic}

To showcase the benefits of dynamic rankings obtained via HawkesRank, we generate data from the SEMHP (Eq.~\eqref{eq:Hawkes}) for $M=10$ types of events.  
The exogenous intensities are drawn from a power law, $\mu_i = i^{-1/2}$, such that nodes with lower indices exhibit systematically higher baseline activity.  

The branching ratio matrix $N$ is constructed in three steps.  
First,  
we sample an unweighted adjacency matrix from the Barab\'asi-Albert ensemble, capturing the mechanism of preferential attachment commonly observed in real-world systems.  
Nodes are added sequentially in the order $i = 1, \ldots, M$, and each newly added node establishes $\eta=5$ outgoing edges, attaching to existing nodes with probability proportional to their current degree.  
Self-loops are included so that diagonal elements $n_{ii} > 0$ represent genuine auto-reinforcement, a defining feature of self-exciting processes.  
Second,  
weights are assigned to the non-zero entries of $N$ by independently sampling them from the distribution on $[0,1]$.
Third,  
the matrix is rescaled to achieve a target branching ratio of $n = 0.6$ by multiplying each non-zero weight by $n / \lambda_\text{max}$.  

We fix the memory kernel decay parameter at $\tau = 1$, so that the effective time scale of fluctuations is approximately $1 / (1 - n)$.  

Given these parameters, we compute the intensities $\lambda_i(t)$ on a fine-grained time grid over $T = 200$ time steps.  
This yields a sequence of instantaneous \emph{ground-truth rankings} $\lambda_i(t)$ that reflect the system's evolving activity: 
At time $t$, event type $i$ is ranked above event type $j$ whenever $\lambda_i(t) > \lambda_j(t)$.
We then assess how well four static centrality benchmarks approximate these dynamic rankings:  
(i) First-moment Hawkes ranking (Eq.~\eqref{eq:first_moment_Hawkes}) using the true $\mu_i$;  
(ii) Katz centrality (Eq.~\eqref{eq:katz0}) with $\beta = \vec{1}$ instead of proper $\mu_i$;  
(iii) Eigenvector centrality, obtained as the limiting case of Katz as $\alpha \to 1 / \lambda_{\max}$;  
(iv) PageRank, which modifies (ii) by replacing the adjacency matrix $A = N$ with its row-normalized counterpart.  

\begin{figure*}[!htb]
	\centering
	\includegraphics[width=\textwidth]{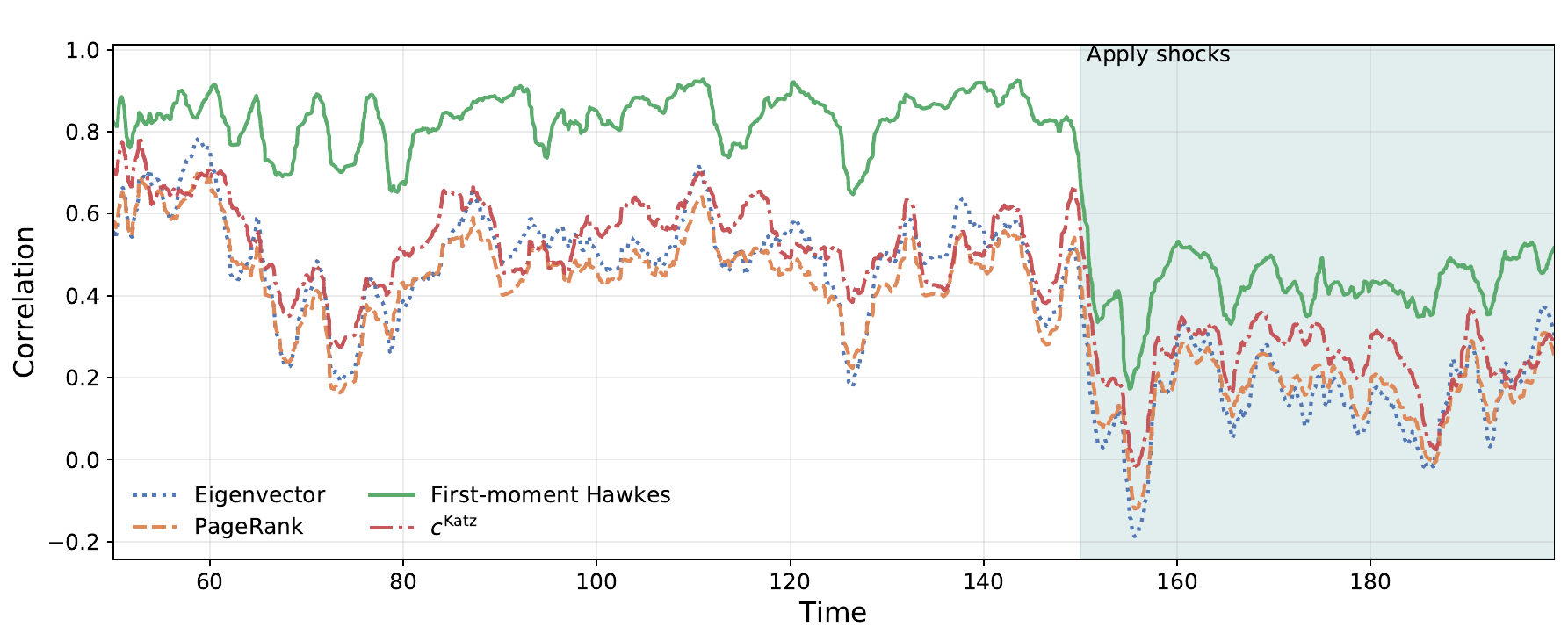}
    \caption{
    For each of four static centrality measures, we compute the Spearman rank correlation between the corresponding static ranking and the ground-truth ranking induced by the instantaneous intensities $\{\lambda_i(t)\}_{i=1}^M$ at each time $t$.  
    Higher correlation values indicate better agreement between the static and dynamic rankings.  
    The vertical dashed line at $t=150$ marks the onset of an exogenous shock, implemented by increasing the smallest baseline intensity $\mu_{10}$ by a factor of $10$.  
    The correlation curves are smoothed using a centered moving average over $50$ time steps to reduce high-frequency stochastic fluctuations in the event-driven rankings.
    }
	\label{fig:simulations}
\end{figure*}

For each benchmark, we compute the Spearman rank correlation between its static ranking and the ground-truth ranking at each time step $t$.
This yields a time series of correlation coefficients for each centrality measure, quantifying their alignment with the system's dynamic behavior over time (Figure~\ref{fig:simulations}).  
We observe that the first-moment Hawkes ranking achieves the best overall performance, followed by conventional Katz centrality with homogeneous exogenous terms.  
This underscores the importance of accounting for exogenous influence, particularly its heterogeneity.

The results are robust to structural parameter choices.  
Qualitatively similar patterns are observed when varying $\eta$ from 1 to 8 and the branching ratio $n$ from 0.3 to 0.9.  
These findings expose a fundamental limitation of static ranking measures.  
Even the best-performing method (the first-moment Hawkes approximation) displays time-varying fluctuations in rank correlation, 
indicating that static approaches are insufficient to capture the dynamics of evolving systems.  

The need for dynamic importance rankings becomes even more evident when exogenous drivers fluctuate in response to shocks or structural changes.  
In social media systems, exogenous attention may shift abruptly following breaking news, viral content, or platform-level interventions, temporarily elevating the activity of specific users or topics.  
In financial markets, external information such as earnings announcements, macroeconomic releases, or regulatory actions can induce sudden changes in trading intensity across assets.  
In neuroscience, external stimuli or experimental perturbations can transiently modulate neuronal firing rates, altering the relative importance of neurons over time.  

To evaluate ranking performance under such conditions, we introduce an exogenous shock to the baseline intensity $\mu$.
Specifically, at time $t = 150$, the smallest exogenous intensity, $\mu_{10}$,  is increased ten-fold and maintained at this elevated level for $50$ time steps.
This intervention perturbs the ordering of baseline intensities and thereby alters the ranking of the simulated intensities.
Figure~\ref{fig:simulations} shows that once the shock occurs, the correlations of all static benchmarks exhibit higher variability and a pronounced decline in agreement with the true rankings.
The first-moment Hawkes ranking provides only limited improvement over the other static methods.
Although it incorporates exogenous contributions, it assumes them to be time-invariant and therefore cannot capture the time-varying external drivers that characterize real-world systems.
This highlights a key limitation of static centrality measures: they cannot adapt to shocks or represent transient dynamics, resulting in less reliable rankings when the system undergoes perturbations.

Since the ground truth ranking is given by HawkesRank based on the full time-dependent intensities, Fig.~\ref{fig:simulations} quantifies how much ranking information is lost when time-varying dynamics are ignored.

\subsection*{Dynamically Ranked Emotions}
\label{sec:emotions}

\begin{figure*}[htb]
    \centering
    \includegraphics[width=0.95\linewidth]{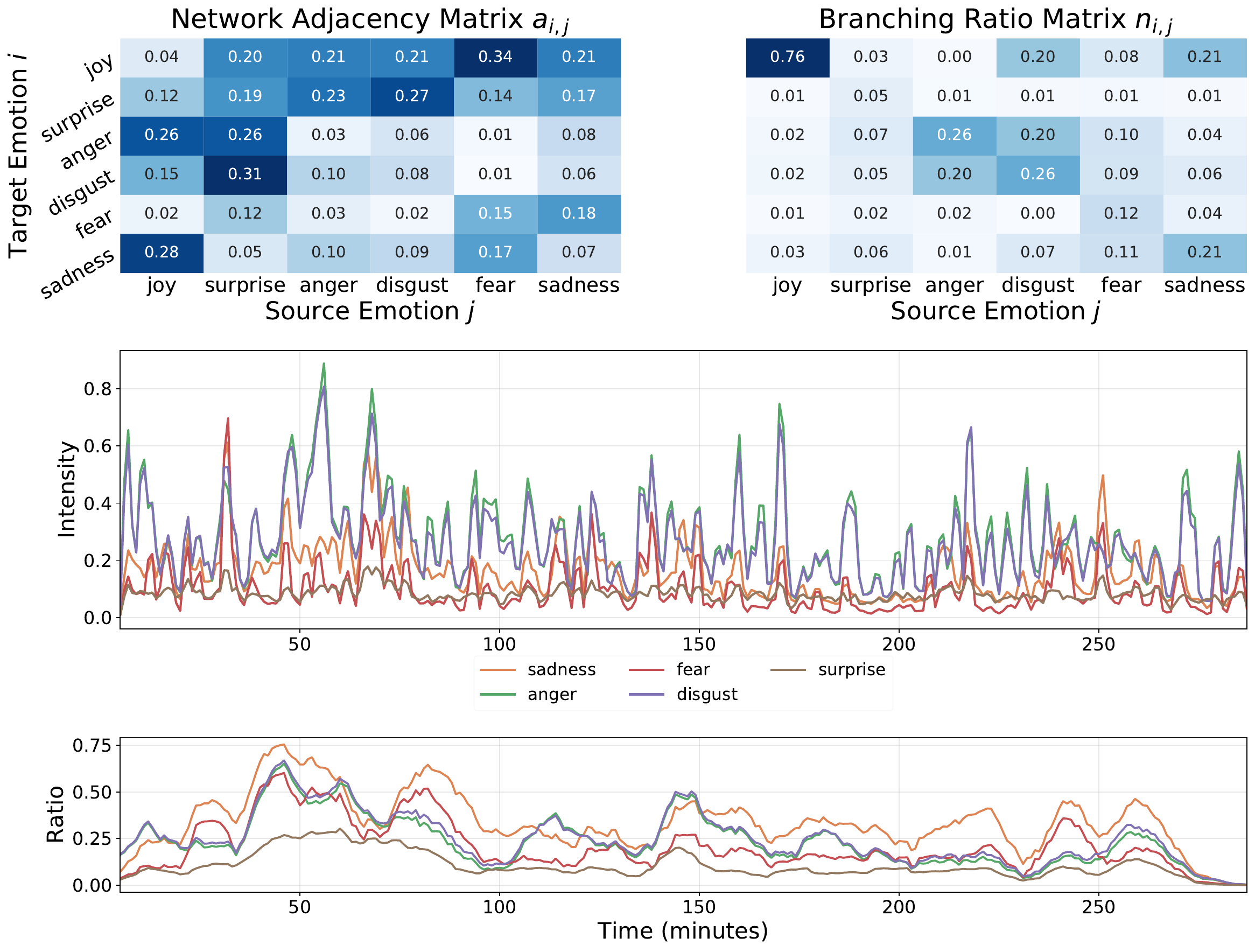}
	\caption{
	Emotion dynamics in a YouTube live-chat video.
	\emph{Top:} 
	Network adjacency matrix (left) and estimated branching ratio matrix $N$ (right), where entry $(i,j)$ quantifies the influence of source emotion $j$ on target emotion $i$.
	\emph{Middle:} 
	Hawkes intensity trajectories $\lambda^{i}(t)$ for five basic emotions across the video. 
	The sixth emotion (\emph{joy}), dominates across emotions and is therefore omitted from plots for visual clarity.
	The intensities are smoothed using a centered 2-time-step moving average.
	\emph{Bottom:} 
	Time-varying ratio of endogenous activity, defined as the proportion of self- and cross-excitation relative to total intensity. 
	The ratios are smoothed using a centered 10-time-step moving average.
	}
    \label{fig:emotion_int_network}
\end{figure*}

Emotional contagion in online communication provides a natural setting for evaluating dynamic ranking methods.
Messages arrive as discrete events, emotional responses arise from both external stimuli and interactions among participants, and the prominence of emotions evolves rapidly over time.
To showcase the benefits of HawkesRank, we thus analyze emotional dynamics in YouTube live-chat discussions.
YouTube Live allows viewers to post messages in real time while watching a streaming video, producing a continuous stream of time-stamped chat messages that reflect audience reactions to the unfolding content.

Following the approach detailed in Ref.~\cite{luo2025social}, emotions expressed in chat messages are categorized according to the widely used six basic emotion framework consisting of joy, surprise, anger, disgust, fear, and sadness \cite{ekman1992}.
Each message is represented as a six-dimensional binary vector indicating the non-exclusive presence of these emotions.
Our objective is to rank the prevalence and influence of these emotions over time, identifying which emotions dominate the conversation at each moment.

A common approach to analyzing emotional influence is to construct a directed network from message activity using lead–lag correlations between emotion time series.
Chat messages are first aggregated into temporal bins of $b$ minutes, yielding time series $x_i(t)$ that record the number of occurrences of emotion $i$ in bin $t$.
For each pair of emotions $(i,j)$, we compute the Pearson correlation between $x_j(t)$ and $x_i(t+\ell b)$, where $\ell$ is an integer lag indicating how many bins separate the two observations.
The adjacency entry $A_{i,j}$ is defined as this lagged correlation, so that an edge $j \rightarrow i$ indicates that fluctuations in emotion $j$ tend to precede fluctuations in emotion $i$ after $\ell$ bins.
Aggregating these relations and normalizing the resulting matrix by its Frobenius norm yields the adjacency matrix shown in the top-left panel of Figure~\ref{fig:emotion_int_network}.

This network construction introduces two free parameters, the bin width $b$ and the lag $\ell$, whose choice is not determined by the data.
In Figure \ref{fig:emotion_int_network}, we have set $b=0.5$ and $\ell=2$.
However, the inferred network structure $A_{ij}$ can vary substantially across plausible parameter values (SI Appendix).
In contrast, the Hawkes framework estimates excitation effects directly from the event timestamps, 
inferring the interaction matrix statistically without requiring arbitrary assumptions to construct a network. 
We therefore model the emotional dynamics using a multivariate Hawkes process.
In this framework, the occurrence of an emotion in a chat message is treated as a discrete event, and emotional propagation is captured through self- and cross-excitation mechanisms.
Each emotion $i$ is characterized by an intensity $\lambda_i(t)$ representing the instantaneous rate at which that emotion appears in the chat stream.
According to Eq.~\eqref{eq:Hawkes}, the branching ratio matrix $N$ captures endogenous excitation whereby past emotional expressions influence future ones.
All parameters are fitted using maximum log-likelihood. 
Additional modeling and estimation details are provided in Ref.~\cite{luo2025social}.

The estimated branching matrix $N$, shown in the top-right panel of Figure~\ref{fig:emotion_int_network}, differs markedly from the adjacency matrix derived from the static network construction, even though both are based on the same underlying data.
In particular, the Hawkes estimation reveals strong self-excitation effects, consistent with prior evidence of emotional contagion in online interactions \citep{kramer2014}.
It also identifies a pronounced bidirectional interaction between anger and disgust that is not captured by the adjacency-based network.
Conversely, the static network misleadingly assigns joy as a dominant source of all other emotions, an interpretation that is behaviorally implausible since expressions of joy rarely induce fear or surprise.
The Hawkes framework disentangles these relationships by separating endogenous excitation from exogenous drivers and by estimating the interaction structure directly from the event data.

Next, we analyze the evolution of the HawkesRank rankings $\lambda_i(t)$ for the six emotions (middle panel of Figure~\ref{fig:emotion_int_network}).
The intensity of the emotion \emph{joy} dominates throughout the observation period and is therefore omitted from the plot to preserve visual resolution.
The remaining five emotions exhibit substantial temporal fluctuations and frequent crossings between trajectories.
These dynamics demonstrate that the relative prominence of emotions changes continuously during the video, highlighting the limitations of static ranking approaches.

Finally, HawkesRank enables a decomposition of total activity into endogenous and exogenous contributions.
Emotions in YouTube Live chats are particularly suitable for such a decomposition because the video itself provides a strongly time-varying exogenous driver.
For example, dramatic or humorous moments in the video may trigger bursts of specific emotional reactions in the chat.
The Hawkes framework therefore distinguishes externally driven responses originating from the video from endogenous amplification among viewers.
The lower panel of Figure~\ref{fig:emotion_int_network} shows that the endogenous share of emotional activity varies considerably over time, indicating that the importance of internal amplification relative to external stimulation is itself dynamic.
Such a decomposition is not available within the static network formulation, yet it reveals fundamental mechanisms underlying emotional dynamics.
More broadly, the framework enables alternative ranking schemes based on total activity, purely exogenous influence, or endogenous amplification.

Together, these findings demonstrate that static centrality measures are insufficient for characterizing and evaluating inherently dynamic systems. 
By collapsing interactions over time, static network approaches obscure temporal heterogeneity and evolving influence patterns.
In contrast, the Hawkes framework explicitly models time-dependent excitation dynamics, yielding instantaneous and adaptive rankings that adjust continuously as the system evolves.

\section*{Discussion \& Conclusions}
\label{sec:conclusion}

We introduced HawkesRank, a dynamic framework for measuring importance in evolving systems based on multivariate Hawkes point processes.
In this formulation, importance is interpreted as the instantaneous event intensity $\lambda_i(t)$, i.e., the expected rate at which events associated with an entity occur.
This gives ranking a direct operational meaning and connects centrality to observable system activity.
Within this framework, widely used measures such as Katz centrality, eigenvector centrality, and PageRank emerge as static mean-field limits, placing classical centrality metrics within a broader event-driven model of influence.

Beyond this unification, HawkesRank offers several conceptual advantages.
It separates exogenous drivers, capturing intrinsic relevance or external shocks, from endogenous amplification generated by network interactions.
At the same time, the branching ratio matrix $N$ provides a statistically inferred interaction structure, replacing heuristic network constructions with influence patterns estimated directly from event data.
Together, these features transform ranking from a purely structural property of networks into a dynamic, interpretable model of event generation that naturally adapts to shocks, feedback, and temporal evolution.

Several directions remain for future work.
On the modeling side, richer excitation structures could be explored by moving beyond simple exponential kernels toward nonparametric or multi-scale memory functions, while also allowing for inhibition, competition, and time-varying interaction matrices $N(t)$.
From a computational perspective, scalable online algorithms with uncertainty quantification will be essential for applying HawkesRank to large systems containing millions of entities.
Theoretical advances are likewise needed to establish identifiability conditions, sample-complexity bounds, and robustness guarantees for recovering $\mu_i(t)$, $N(t)$, and the associated rankings from event data.

Expanding empirical applications will further clarify the practical value of the framework.
In financial markets, HawkesRank may help detect speculative bubbles by identifying when endogenous amplification dominates fundamental drivers.
In social media, it can distinguish genuine influence from artificially boosted visibility.
In science, it may help separate intellectual impact from citation manipulation.
In public health, it can help disentangle baseline transmission from transient spikes caused by super-spreading events.

By grounding ranking in event intensities and explicitly modeling memory and feedback, HawkesRank provides an interpretable, data-driven, and adaptive notion of importance.
We view this event-based perspective as a foundation for the next generation of ranking systems capable of explaining, predicting, and guiding decisions in complex, rapidly evolving environments.

\bibliographystyle{elsarticle-num}
\bibliography{bibliography}

\section*{Methods}
\label{sec:methods}

\subsection*{Classical Centrality Measures}
\label{sec:methods_centrality}

To benchmark our approach, we consider three widely used centrality measures in network science: eigenvector centrality, Katz centrality, and PageRank.

\emph{Eigenvector centrality.}
Eigenvector centrality is based on the principle that a node is important if it is connected to other important nodes.
For a network with adjacency matrix $A_{ij}$, the centrality score $c_i$ satisfies
\begin{equation}
c_i^{(\mathrm{eig})} = \frac{1}{\lambda} \sum_{j=1}^{N} A_{ij} c_j^{(\mathrm{eig})},
\end{equation}
which in vector form corresponds to the eigenvalue problem
\begin{equation}
Ac = \lambda c,
\end{equation}
where $\lambda$ is the largest eigenvalue of $A$.

\emph{Katz centrality.}
Katz centrality extends eigenvector centrality by incorporating both direct and indirect connections.
Let $(A^k)_{ij}$ denote the number of walks of length $k$ from node $j$ to node $i$, and let $\alpha$ be a damping factor that discounts longer paths.
The Katz centrality score is defined as
\begin{equation}
c_i^{(\mathrm{katz})} = \sum_{k=1}^{\infty} \sum_{j=1}^{N} \alpha^k (A^k)_{ji}.
\end{equation}
Equivalently, it can be written in recursive form as
\begin{equation} \label{apx:Katz_recursive}
c_i^{(\mathrm{katz})} = \alpha \sum_{j=1}^{N} A_{ij} c_j^{(\mathrm{katz})} + \beta_i,
\end{equation}
where $\beta_i$ represents exogenous contributions to node importance.
Convergence requires $\alpha < 1/\lambda$, where $\lambda$ is the largest eigenvalue of $A$.

\emph{PageRank.}
PageRank is a variant of eigenvector centrality designed for directed networks such as the web \cite{brin1998}.
The PageRank score of node $i$ is defined as
\begin{equation}
PR(i) = (1-d)\frac{1}{N} + d \sum_{j \in M(i)} \frac{1}{L(j)} PR(j),
\end{equation}
where $M(i)$ is the set of nodes linking to $i$, $L(j)$ is the number of outgoing links from node $j$, and $d$ is a damping factor typically set to $0.85$.
The first term represents random jumps across the network, ensuring that every node has a nonzero probability of being visited.

Additional discussion and mathematical details are provided in the SI Appendix.

\subsection*{Multivariate Hawkes Process}
\label{sec:methods_hawkes}

Consider $M$ event types labeled $i=1,\ldots,M$, where each type corresponds to events occurring on a particular object (e.g., a webpage, user, or asset).
The occurrence of events is characterized by an intensity function $\lambda_i(t)$, which specifies the instantaneous rate at which events of type $i$ occur at time $t$.
The probability of observing an event of type $i$ in a short interval $[t,t+dt)$ is $\lambda_i(t)\,dt$.
The multivariate Hawkes process is defined as
\begin{equation}
\lambda_i(t)
=
\mu_i(t)
+
\sum_{j=1}^{M}
\int_{-\infty}^{t}
f_{j,i}\,\phi_{j,i}(t-s)\,dN_j(s),
\label{eq:Hawkes_general_methods}
\end{equation}
where $N_j(s)$ denotes the counting process of events of type $j$ up to time $s$.

The first term $\mu_i(t)$ represents the exogenous or background rate of events, capturing activity driven by external factors or intrinsic attractiveness.
The second term describes endogenous excitation, whereby past events increase the likelihood of future events.

Two elements determine the strength and timing of this excitation.
First, the coefficient $f_{j,i}$ represents the expected number of type-$i$ events triggered by a single event of type $j$, capturing the strength of cross-excitation between event types.
Second, the kernel $\phi_{j,i}(t)$ describes how the influence of past events decays over time.
In our implementation we use an exponential kernel
$
\phi_{j,i}(t) = \left. \exp \left( -t/\tau_{j,i} \right) \right/ \tau_{j,i}, 
$
which defines a characteristic memory timescale $\tau_{j,i}$.

The expected number of first-generation events of type $i$ generated by an event of type $j$ is given by
$
n_{j,i} = \mathbb{E}[f_{j,i}],
$
and the collection of these coefficients forms the branching ratio matrix
$
N = (n_{j,i}).
$
The spectral radius of $N$, denoted $n$, determines the stability of the process \cite{DaleyVere2008}.

For analytical tractability, we consider the commonly used mean-field formulation in which fertilities are fixed at their expectations, $f_{j,i}=n_{j,i}$.
Under this assumption the intensity reduces to
\begin{equation}
\lambda_i(t)
=
\mu_i(t)
+
\sum_{j=1}^{M}
\int_{-\infty}^{t}
n_{j,i}\,\phi(t-s)\,dN_j(s),
\label{fdhbwgvq}
\end{equation}
which corresponds to the formulation used in the Results section.

\subsection*{First-Moment Representation and Connection to Classical Centrality}
\label{sec:methods_firstmoment}

The Hawkes process formulation admits a direct connection to classical network centrality measures through its first moment.
Taking the expectation of the intensity in Eq.~\eqref{fdhbwgvq} over all stochastic realizations yields the mean-field equation
\begin{equation}
\mathbb{E}[\lambda_i](t)
=
\mu_i(t)
+
\sum_{j=1}^{M}
\int_{-\infty}^{t}
\mathbb{E}[\lambda_j](s)\, n_{j,i}\,\phi(t-s)\,ds.
\label{eq:first_moment_methods}
\end{equation}
This equation describes how the expected activity of each entity results from both exogenous inputs and cascades generated through the interaction matrix $N=(n_{j,i})$.

Two limiting regimes connect this formulation to classical centrality metrics.
First, when the memory kernel becomes infinitely short ($\tau \rightarrow 0$), the kernel approaches a Dirac delta function and Eq.~\eqref{eq:first_moment_methods} reduces to
\begin{equation}
\mathbb{E}[\lambda_i](t)
=
\mu_i(t)
+
\sum_{j=1}^{M}
n_{j,i}\,\mathbb{E}[\lambda_j](t).
\end{equation}
In vector form this yields the stationary solution
\begin{equation}
\mathbb{E}[\Lambda] = (I-N)^{-1}\mu.
\label{eq:firstmoment_stationary}
\end{equation}

Equation~\eqref{eq:firstmoment_stationary} is equivalent to the recursive form
\begin{equation}
\mathbb{E}[\Lambda] = \mu + N\,\mathbb{E}[\Lambda],
\end{equation}
which directly parallels the Katz formulation from Eq. \eqref{apx:Katz_recursive}.
Identifying $\mu=\vec{\beta}$ and $N=\alpha A$ therefore shows that the stationary first moment of the Hawkes process recovers Katz centrality.

The same stationary expression also arises in the long-time limit of the Hawkes process when the exogenous intensity $\mu(t)$ is constant.
In that case, the convolution term in Eq.~\eqref{eq:first_moment_methods} converges to a steady value and the expectation becomes time-independent,
yielding the fixed-point solution Eq.~\eqref{eq:firstmoment_stationary}.
Full derivations are provided in the SI Appendix.

\balance
\clearpage
\appendix
\onecolumn

\section{Review of Centrality Measures}
\label{apx:review}

This appendix reviews the standard definitions and limitations of three widely used centrality metrics:
eigenvector centrality, Katz centrality, and PageRank, all of which are relevant to our work and closely related to one another.  

\subsection{Definitions of Centrality Measures}

We begin by defining each centrality measure in its classical form, highlighting key mathematical properties and connections.

\textit{Eigenvector Centrality.}  
Eigenvector centrality is based on the self-consistent principle that a node is important if it is connected to other important nodes.  
For a network with adjacency matrix $A_{ij}$, the eigenvector centrality of node $i$ is given by  
\begin{equation}
c_i^{\text{(eig)}} = \frac{1}{\lambda} \sum_{j=1}^{N} A_{ij} c_j^{\text{(eig)}},
\end{equation}
where $\lambda$ is the largest eigenvalue of the matrix $A$.  
In vector form, this corresponds to the standard eigenvector equation:
\begin{equation}
Ac = \lambda c.
\end{equation}
The score $c_i$ is thus proportional to the sum of the centralities of its neighbors.  
This centrality is computationally efficient and forms the foundation for more refined variants, including Katz centrality and the PageRank algorithm.

\textit{Katz Centrality.}  
Katz centrality generalizes eigenvector centrality by accounting not only for immediate neighbors but also for indirect connections of all lengths.  
Let $(A^k)_{ij}$ denote the number of walks of length $k$ from node $j$ to node $i$, and let $\alpha \in (0, 1)$ be a damping factor that attenuates longer paths.  
Then the Katz centrality of node $i$ is defined as:
\begin{equation}
c_{i}^{(\text{katz})} = \sum_{k=1}^{\infty} \sum_{j=1}^{N} \alpha^k (A^k)_{ji}.
\label{eq:katz_1}
\end{equation}
This represents a discounted sum over all walks leading to node $i$, assigning greater weight to shorter paths.  
Nodes that are easier to reach via many short paths receive higher centrality scores.

Defining the identity matrix of size $N$ as $I$, and the all-ones vector as $\vec{I}$, this expression can be rewritten compactly as:
\begin{equation}
c^{\text{katz}} = \left( \sum_{k=1}^{\infty} (\alpha A^T)^k \right) \vec{I}.
\label{eq:katz_2}
\end{equation}
Using the geometric series identity, this becomes:
\begin{equation}
c^{\text{katz}} = \left( (I - \alpha A^T)^{-1} - I \right) \vec{I}.
\end{equation}
Multiplying both sides by $(I - \alpha A^T)$ yields:
\begin{subequations}
\begin{align}
c^{\text{katz}} (I - \alpha A^T) &= \vec{I} - (I - \alpha A^T)\vec{I} \\
&= \alpha A^T \vec{I}.
\end{align}
\end{subequations}
Equivalently,
\begin{subequations}
\begin{align}
c^{\text{katz}} &= \alpha A^T \vec{I} + \alpha A^T c^{\text{katz}} \\
&= \alpha A^T (c^{\text{katz}} + \vec{I}),
\end{align}
\end{subequations}
which can be written component-wise as:
\begin{equation}
c_i^{\text{(katz)}} = \alpha \sum_{j=1}^{N} A_{ij} (c_j^{\text{(katz)}} + 1).
\end{equation}
To ensure that the matrix $(I - \alpha A^T)$ is invertible, we require $\alpha < 1/\lambda$, where $\lambda$ is the largest eigenvalue of $A$.  
This condition ensures sufficient discounting of long paths to guarantee convergence.

The definition in Equations~\eqref{eq:katz_1} and \eqref{eq:katz_2} assumes equal influence from all source nodes by using the uniform vector $\vec{I}$.  
A natural generalization replaces $\vec{I}$ with a vector $\vec{\beta}$, which allows for heterogeneous exogenous contributions:
\begin{equation}
c_i^{\text{(katz)}} = \alpha \sum_{j=1}^{N} A_{ij} c_j^{\text{(katz)}} + \beta_i.
\label{eq:katz}
\end{equation}
As $\vec{\beta} \to 0$ and $\alpha \to 1/\lambda$, Katz centrality converges to eigenvector centrality.

\textit{PageRank.}  
The PageRank algorithm, originally developed to rank webpages, is a variant of eigenvector centrality adapted to directed hyperlink networks \cite{brin1998}.  
It assigns higher scores to nodes that are linked by other high-ranking nodes, while incorporating a damping factor to prevent rank sinks.

The PageRank of node $i$ is defined by the recursive equation:
\begin{equation}
PR(i) = (1 - d)\frac{1}{N} + d \sum_{j \in M(i)} \frac{1}{L(j)} PR(j),
\label{eq:Page-rank}
\end{equation}
where $M(i)$ is the set of nodes linking to $i$, and $L(j)$ is the number of outbound links from node $j$.  
The damping factor $d \in (0, 1)$, typically set to $0.85$, represents the probability that a user follows a link rather than jumping to a random page.

The first term ensures that each page has a nonzero probability of being visited, mitigating the effects of disconnected or cyclic subgraphs.  
When $d = 1$, the formulation reduces to pure eigenvector centrality.  
PageRank has been widely adopted in domains beyond web search, including citation networks, social systems, protein interaction networks, and sports rankings \cite{gleich2015}.

\subsection{Limitations of Centrality Measures in Network Analysis}

Classical centrality measures exhibit several critical limitations that constrain their applicability to real-world, dynamic systems.
These include divergence from fundamental values, static formulations that ignore temporal evolution, and reliance on arbitrary or ad hoc parameters. 
Below, we synthesize these issues and review related extensions and partial remedies.

First, centrality-based rankings can deviate significantly from intrinsic value. 
Because these methods amplify network-based visibility, they are susceptible to manipulation and self-reinforcing distortions. 
Examples include search engine optimisation (SEO) and link farming, the artificial boosting of social media influencers, and the formation of echo chambers where visibility compounds independently of quality. 
Like financial markets, centrality systems can exhibit bubbles-phases where perceived importance diverges from fundamental worth, 
fueled by endogenous feedback rather than intrinsic merit.
Valuable but less popular nodes are often eclipsed by highly connected but lower-quality content, reflecting an emphasis on influence over substance.

Second, most classical metrics are static. 
They capture the state of a network at a single moment and fail to account for ongoing changes in structure or context. 
In reality, node importance is highly dynamic: 
news outlets gain prominence during major events, influencers rise and fall with viral content, and geopolitical or market shifts suddenly reconfigure the strategic relevance of institutions or infrastructure. 
Static metrics overlook such fluctuations and risk producing rankings that are outdated, misleading, or blind to emerging threats and opportunities. 
Dynamic frameworks, such as the endo-exo ranking approach proposed here, address this by modeling both exogenous shocks and endogenous reinforcement over time.

Third, widely used metrics depend on arbitrary damping factors or weights. 
For instance, PageRank typically uses a damping factor of 0.85, chosen heuristically rather than derived from first principles \citep{bonacich1987power}. 
More broadly, external information is often incorporated through uniform or ad hoc weight vectors, offering little interpretability or adaptability. 
Adjusting these parameters to fit new contexts requires costly manual tuning and trial-and-error. 
Moreover, static weights cannot respond to shifting dynamics or new data without re-specification.

Several extensions have been proposed to address these shortcomings. 
Time-aware adaptations such as Temporal PageRank \citep{rozenshtein2016temporal, kim2012temporal} and snapshot-aggregation methods \citep{grindrod2011communicability, mendonca2021approximating} introduce some temporal responsiveness but remain fundamentally based on static formulations. 
Continuous-time ODE models \citep{grindrod2014} generalize Katz centrality while avoiding discretization but lack stochastic dynamics, event-level resolution, or an explicit separation between endogenous and exogenous effects.

Recent data-driven models attempt to learn dynamic influence scores directly from observed cascades or metadata using machine learning or statistical inference \citep{grando2018machine, li2023centrality, stolarski2024identifying}. 
While flexible and context-aware, such methods often require large annotated datasets, are computationally intensive, and lack interpretability. Moreover, they typically optimize for predictive performance rather than theoretical consistency or explanatory clarity.

Taken together, these observations point to a fundamental limitation of existing centrality metrics: 
they rely heavily on structural reinforcement while neglecting exogenous drivers and intrinsic value. As a result, they risk misrepresenting influence, misallocating attention and resources, and reinforcing systemic vulnerabilities.
A principled alternative requires a shift toward dynamic, event-driven, and interpretable ranking systems that jointly model memory, network reinforcement, and exogenous signals. These goals are addressed by the HawkesRank framework developed in this work.

\section{Stationary First Moment of the SEMHP Recovers Katz Centrality}
\label{apx:recover_Katz}

In this appendix, we present the theoretical result that the stationary first moment of the Hawkes process recovers Katz centrality (Eq.~\eqref{eq:katz}) as a special case. This mapping underscores the Hawkes process as a unifying framework for ranking in complex systems, while at the same time revealing the inherent limitations of traditional centrality measures.

\subsection{General Expression of the Multivariate Hawkes Process}
\label{apx:general_hawkes}

We start by presenting the general form for the Multivariate Hawkes Process from which the simplified version 
(\ref{eq:Hawkes}) presented in the main article is derived.

First, let us note that the standard mathematical formulation in the language of point process replaces expression (\ref{eq:Hawkes}) by
\begin{equation}
	\lambda_i(t)
	=  \underbrace{\mu_i(t)}_{\text{exo}} + \underbrace{ \sum_{j =1}^M \nt{-\infty}{t}{N_j(s)} n_{j,i} \phi\left( t-s \right) }_{\text{endo}}.
	\label{eq:Hawkes222}
\end{equation} 
In this formula, the variable $N_j(s)$ denotes the counting process of type-$j$ events up to time $s$, and $\mathrm{d}N_j(s)$ represents its infinitesimal increment at time $s$.  
The stochastic integral of a function $g(t)$ with respect to $N$ is defined as
\begin{equation}
\int_{-\infty}^t g(t)\,\mathrm{d}N(s) := \sum_{j\,:\,t_j \leq t} g(t_j),
\label{dhyh3h}
\end{equation}
where $\{t_j\}$ is the set of event timestamps recorded by the counting process $N_j(s)$. 
Using this definition of the stochastic integral (\ref{dhyh3h}) recovers (\ref{eq:Hawkes}) from (\ref{eq:Hawkes222}).

We now generalise this formulation. We again consider $M$ different event types, labeled $i = 1, \ldots, M$, where each type corresponds to events occurring on a specific object (for example, a user in a social network or a financial asset). For each type $i$, the process specifies an intensity function $\lambda_i(t)$ that describes the instantaneous rate at which events of type $i$ are expected to occur at time $t$. Intuitively, $\lambda_i(t)$ plays the role of a time-varying hazard rate: the probability of observing an event of type $i$ in a short interval $[t, t+dt)$ is equal to $\lambda_i(t),dt$.
Formally, the intensity is given by
\begin{equation}
\lambda_i(t)
= \underbrace{\mu_i(t)}_{\text{exo (background rate)}}
+ \underbrace{ \sum_{j =1}^M \int_{-\infty}^{t} f_{j,i} ~ \phi_{j,i}(t-s) ~ dN_j(s)}_{\text{endo (self- and cross-excitation)}} .
\label{eq:Hawkes_general}
\end{equation}
This expression has two parts:

\begin{itemize}

\item
\emph{Exogenous component.}  

Events may arise independently of prior occurrences, driven by external influences or intrinsic attractiveness.  
In the context of website traffic, $\mu_i(t)$ captures all forces that direct attention toward page $i$ without relying on internal network dynamics.  
These forces include media coverage, advertising, social buzz, and the intrinsic appeal or relevance of the content itself.  
A webpage may therefore attract visits through reputation, topical relevance, or external exposure that amplifies its visibility independently of link-based propagation.

\item
\emph{Endogenous component.}  

Past events increase the likelihood of future events through self- and cross-excitation mechanisms.  
An event of type $j$ occurring at time $s$ contributes to the intensity of type $i$ at a later time $t$ through two distinct factors:

\begin{enumerate}

\item
The fertility coefficient $f_{j,i}$ specifies the expected number of type-$i$ offspring events directly caused by a single type-$j$ parent event.  
More precisely, $f_{j,i}$ denotes a stochastic fertility drawn from a probability distribution and represents the average number of type-$i$ events triggered by a type-$j$ event.  
This construction gives rise to a doubly stochastic process.  
First, when an event of type $j$ occurs at time $s$, it is associated with a randomly drawn fertility vector $(f_{j,i})_i$.  
Second, conditional on this fertility vector, the actual number of daughter events of type $i$ generated by the parent event of type $j$ is random and follows a Poisson distribution with mean $f_{j,i}$.  
This formulation captures both idiosyncratic event-level heterogeneity, through randomness in fertilities, and stochasticity in event propagation, through Poisson offspring generation.  

\item
The kernel $\phi_{j,i}(t-s)$ describes how the influence of a past event decays over time.  
It models the contribution of events of type $j$ occurring prior to time $t$ to the intensity of type $i$ at time $t$.  
We assume an exponential decay of the form $\phi_{j,i}(t) = \frac{1}{\tau_{j,i}} e^{-t / \tau_{j,i}}$, where $\tau_{j,i}$ is a memory parameter.  
This specification can be generalized to other temporal profiles, such as power-law decay, to capture long-memory effects.  
The memory kernel $\phi_{j,i}(t-s)$ defines the waiting-time distribution for an event of type $i$ to occur at time $t$ given that an event of type $j$ occurred at time $s$.  
Each kernel is normalized such that $\int_0^{\infty} \phi_{j,i}(u)\,\mathrm{d}u = 1$.  
For simplicity, and without loss of generality for our main results, we assume that all memory parameters $\tau_{j,i}$ are identical and equal to a common value $\tau$.  
Under this assumption, the kernel reduces to $\phi_{j,i}(t-s) = \phi(t-s) = \frac{1}{\tau} e^{-(t-s)/\tau}$.

\end{enumerate}
\end{itemize}

In short, the Hawkes process captures the idea that events can arise both from a background rate of activity and from cascades triggered by earlier events. This allows us to model systems where influence and attention spread dynamically across different objects.

One can define the average number of first-generation daughter events of type $i$ triggered by a mother event of type $j$,
\begin{equation}
    n_{j,i} = \mathbb{E}[f_{j,i}],
\end{equation}
where the expectation is taken over all possible draws of the stochastic fertilities. This defines the ``branching ratio'' from the ``mother event'' of type $j$
to ``daughter events'' of type $i$. The collection of these branching ratios forms the branching ratio matrix
\begin{equation}
    N = (n_{j,i})   ~~~{\rm or~ equivalently}~~~N^T = (n_{i,j})
    \label{dhtyh3wybn2}
\end{equation}
Its spectral radius, denoted $n$, corresponds to the largest eigenvalue of $N$ and determines the stability of the dynamics \cite{DaleyVere2008}.

To obtain the  simplified version presented in the main article, 
we adopt the following assumption: the fertilities $f_{j,i}$ are no longer random and are fixed at their expectations $n_{j,i}$. This approximation removes randomness in the mothers' fertilities, so that an event of type $j$ generates, on average, $n_{j,i}$ first-generation events of type $i$.
This simplification yields the multivariate Hawkes process in the form given by expression (\ref{eq:Hawkes222}), which is equivalent to 
expression (\ref{eq:Hawkes}), as already mentioned.

\subsection{First Moment of the Hawkes Process}
\label{sec:Hawkes_mapping}

Equation \eqref{eq:Hawkes} defines the intensity function $\lambda_i(t)$ of a Hawkes process for process $i$ at time $t$.
We show that the stationary first moment of the Hawkes process recovers Katz centrality (Eq.~\eqref{eq:katz}) under either of the following two conditions:  
1) when the memory parameter tends to zero, or  
2) in the asymptotic limit as time approaches infinity.

The Hawkes process is a stochastic point process.
Calculating the first moment of the Hawkes process amounts to taking the expectation of expression \eqref{eq:Hawkes} with respect to all possible stochastic realizations of the process, which gives:
\begin{subequations}
\begin{align}
        \mathbb{E}[\lambda_i](t) 
        &= \mathbb{E}[\mu_i(t)] + \mathbb{E}\left[\sum_{j =1}^M \nt{-\infty}{t}{N_j(s)} n_{j,i} \phi \left( t-s \right)\right] \\
        &= \mu_i(t) + \sum_{j =1}^M \int_{-\infty}^{t} \mathbb{E}[\mathrm{d}{N_j(s)}] n_{j,i}\phi ( t-s ) \\
        &= \mu_i(t) + \sum_{j =1}^M \int_{-\infty}^{t} \mathrm{d}s ~ \mathbb{E}[\lambda_j](s) n_{j,i} \phi( t-s ). 
        \label{eq:first_moment_general}
\end{align}
\end{subequations}

\subsubsection{Limit of Infinitely Small Memory Parameter}
When the memory parameter $\tau$ approaches 0, the function $\phi(t)$ approaches a Dirac functions $\delta$, which gives: 
\begin{subequations}
\begin{align}
        \mathbb{E}[\lambda_i](t) 
        &= \mu_i(t) + \sum_{j =1}^M \int_{-\infty}^{t} \mathrm{d}s ~ \mathbb{E}[\lambda_j](s) n_{j,i} \delta( t-s ) \\
        &= \mu_i(t) + \sum_{j =1}^M \mathbb{E}[\lambda_j](t)n_{j,i}.  
        \label{eq:first_moment_Hawkes1}
\end{align}
\end{subequations}
By taking $\mu_i = \vec{\beta}$ and $n_{j,i}= \alpha A_{j,i}$, the stationary first moment of the Hawkes process recovers the Katz centrality in Equation \eqref{eq:katz}.
The first moment Hawkes in the univariate case is solution of 
\begin{equation}
        \mathbb{E}[\lambda](t) = \mu(t) + n\mathbb{E}[\lambda](t),
        \label{eq:first_moment_Hawkes_uni}
\end{equation}
which yields 
\begin{equation}
        \mathbb{E}[\lambda](t) = {\mu(t) \over 1-n} ~.
        \label{eq:first_moment_Hawkes_uqerg1revni}
\end{equation}
As the branching ratio $n$ approaches the critical value $1$, the intensity diverges due to runaway self-excitation.  
The resulting $1/(1-n)$ dependence has a straightforward interpretation: starting from a single mother event, on average it produces $n$ daughters, each of which generates $n$ granddaughters, and so on.  
The total expected number of events is therefore the geometric series  
$1 + n + n^2 + n^3 + \cdots = \frac{1}{1-n}$, 
which diverges as $n \to 1$.

\subsubsection{Asymptotic Long-Time Limit with Constant Exo-Term}
Here, we show that the stationary first moment of the Hawkes process recovers the Katz centrality under the condition
of taking the long-term time limit.

The Laplace transform of a function $f(x)$ is defined as $\mathcal{L}\{f\}(\beta) = \tilde{f}(\beta) = \nt{0}{\infty}{t} f(t)e^{-\beta t}$.
Because the Laplace transform of a convolution is given by a multiplication in the Laplace space, the Laplace transform of Equation \eqref{eq:first_moment_general} is given by:
\begin{equation}
        \mathbb{E}[\tilde{\lambda_i}](\beta) =  \tilde{\mu_i}(\beta) + \sum_{j =1}^M n_{i,j} \mathbb{E}[\tilde{\lambda_j}(\beta)] \tilde{\phi_{i,j}}(\beta).
        \label{eq:laplace_hawkes}
\end{equation}
For a univariate Hawkes process where the number of event types is $M = 1$, we can immediately solve for the first order moment as: 
\begin{subequations}
\begin{align}
        \mathbb{E}[\tilde{\lambda}](\beta)
        &= \tilde{\mu_i}(\beta) + n \mathbb{E}[\tilde{\lambda}(\beta)] \tilde{\phi}(\beta) \\
        &= \frac{\tilde{\mu_i}(\beta)}{1-n \tilde{\phi}(\beta) }.
\end{align}
\end{subequations}
The Laplace transform of a constant $a$ is given by $\tilde{a}(\beta) = \frac{a}{\beta}$.
When the exogenous intensity is time-invariant, $\mu(t) = \mu$, the Laplace transform of $\mu$ is $\tilde{\mu}(\beta) = \frac{\mu}{\beta}$.
Substituting in $\tilde{\mu}(\beta)$ gives
\begin{equation}
        \mathbb{E}[\tilde{\lambda}](\beta) = \frac{\frac{\mu}{\beta}}{1-n \tilde{\phi}(\beta)}.
\end{equation}
The Laplace transform of the exponential kernel $\phi(t) = {\frac{1}{\tau}} e^{-\frac{1}{\tau} t}$ is given by
\begin{equation}
        \tilde{\phi}(\beta) = \frac{1}{\tau \beta + 1}.
\end{equation}
The Taylor expansion of $\tilde{\phi}(\beta)$ in powers of $\beta$ is given by
\begin{equation}
        \tilde{\phi}(\beta) = 1 - \tau \beta + \mathcal{O}(\beta^2).
        \label{eq:taylor}
\end{equation}
Because the long-term limit $t \to \infty$ in the time domain corresponds the limit $\beta \to 0$
in the conjugate Laplace domain, after substituting in $\tilde{\phi}(\beta)$, we obtain the stationary long-term solution as,
\begin{equation}
        \mathbb{E}[\tilde{\lambda}](\beta)
        = \frac{\mu}{1-n}\frac{1}{\beta}\left(1+\frac{n}{1-n} \tau \beta\right)^{-1} 
        = \frac{\mu}{1-n}\frac{1}{\beta}.
\end{equation}
This recovers the shape of a constant inverse Laplace transform, and hence by taking the inverse Laplace transform:
\begin{equation}
        \mathbb{E}[\lambda] = \frac{\mu}{1-n},
\end{equation}
which recovers Equation \eqref{eq:first_moment_Hawkes_uqerg1revni}.

The derivation can be extended to a multivariate Hawkes process where the number of event types $M > 1$.
We denote $Z(\beta) \equiv N \tilde{\phi(\beta)}$, where ${N}$ is the matrix of branching ratios $n_{i,j}$ defined
in equation (\ref{dhtyh3wybn2}). We also define
\begin{equation}
\tilde{\Lambda}(\beta) =
\begin{pmatrix}
\tilde{\lambda_1}(\beta) \\
\vdots \\
\tilde{\lambda_M}(\beta)
\end{pmatrix}  ~~~~~~~~~~~~~
\tilde{\mu}(\beta) =
\begin{pmatrix}
\tilde{\mu_1}(\beta) \\
\vdots \\
\tilde{\mu_M}(\beta)
\end{pmatrix}.
\end{equation}
Equation \eqref{eq:laplace_hawkes} is thus written as:
\begin{subequations}
\begin{align}
        \mathbb{E}[\tilde{\Lambda}](\beta)
        &= \tilde{\mu}(\beta) + \tilde{Z}(\beta) \mathbb{E}[\tilde{\Lambda}](\beta) \\
        &= \tilde{\mu}(\beta)(1 - \tilde{Z}(\beta))^{-1}.
\end{align}
\end{subequations}
The Taylor expansion of $\tilde{\phi}_(\beta)$ as shown in equation \eqref{eq:taylor} generalizes to vectors in a straightforward way.
Substituting in $\tilde{\phi}(\beta)$ obtains
\begin{equation}
        \tilde{Z}(\beta) = N (1 - \tau \beta + \mathcal{O}(\beta^2)).
\end{equation}
As $\beta \to 0$, $\tilde{Z}(\beta) = N$, and
\begin{equation}
\mathbb{E}[\tilde{\Lambda}](\beta) = [1 - N]^{-1}
\begin{pmatrix}
\frac{\tilde{\mu}_1}{\beta} \\
\vdots \\
\frac{\tilde{\mu}_M}{\beta}
\end{pmatrix}.
\end{equation}
The inverse Laplace transform of $\mathbb{E}[\tilde{\Lambda}](\beta)$ is given by
\begin{equation}
\mathbb{E}[\Lambda](t) = [1 - N]^{-1}
\begin{pmatrix}
\mu_1 \\
\vdots \\
\mu_M
\end{pmatrix},
\label{trhjt1ght1}
\end{equation}
which is the multivariate generalisation of expression (\ref{eq:first_moment_Hawkes_uqerg1revni}).

This shows that by taking $\mu = \vec{\beta}$ and $n_{i, j} = A_{ij}$, the stationary first moment of the Hawkes process recovers the Katz centrality in Equation \eqref{eq:katz} with $\alpha = 1$.
Equation \eqref{eq:first_moment_Hawkes1} also recovers the Eigenvector centrality by taking $\mu \to 0$ and $n_{i, j} = \frac{A_{ij}}{n}$.
This reveals a key assumption of the Eigenvector centrality that the system is operating at criticality where the intensities are driven purely by endogenous impacts.
In addition, Equation \eqref{eq:first_moment_Hawkes1} recovers the PageRank algorithm in Equation \eqref{eq:Page-rank} by taking $\mu = \frac{1-d}{N}$ and $n_{i, j} = \frac{d}{L(j)}$.
The PageRank algorithm is a simplification of the first moment Hawkes assuming a homogeneous constant exogenous intensity, homogeneity in the branching ratio matrix, and a stationary process with no time dependency.

In the framework of the Hawkes process, the node centrality score can be interpreted as the predicted probability of observable events.
We show through the mapping that several centrality measures can be recovered by the first moment Hawkes by making various simplifications, which translates into assumptions imposed on the network systems.
This mapping provides an overarching framework to examine and compare different centrality measures and their applicability to social networks.

\subsubsection{Asymptotic Long-Time Limit with Non-Constant Exo-Term}
In the previous subsection, the Katz mapping was derived under the assumption that the exogenous term is
time-invariant, $\mu_i(t) \equiv \mu_i$.  
We now consider the case where $\mu(t)$ varies in time.

Taking Laplace transforms of the first-moment equation as in \eqref{eq:laplace_hawkes}, the mapping from
$\mu(t)$ to $\mathbb{E}[\lambda(t)]$ is given in the frequency domain by
\begin{equation}
\mathbb{E}[\tilde{\Lambda}](\beta) = \left( I - N \,\tilde{\phi}(\beta) \right)^{-1} \tilde{\mu}(\beta),
\label{eq:transfer_nonconstant}
\end{equation}
so that $\mu(t)$ is filtered by a matrix-valued transfer function
\begin{equation}
H(\beta) := \left( I - N \,\tilde{\phi}(\beta) \right)^{-1}.
\end{equation}

For exponential kernels, $\tilde{\phi}(\beta) = (1 + \tau\beta)^{-1}$, and in the low-frequency limit
$\beta \to 0$, one has $\tilde{\phi}(\beta) \to 1$, recovering the Katz factor
\begin{equation}
H(\beta) \xrightarrow[\beta \to 0]{} (I-N)^{-1}.
\end{equation}
Thus, slow components of $\mu(t)$ (or long-time averages) are amplified exactly as in the Katz mapping.

For finite frequencies, however, the gain becomes frequency-dependent, 
\begin{equation}
H(\beta) = \left(I - \frac{N}{1+\tau\beta} \right)^{-1}
\end{equation}
which acts as a matrix-valued low-pass filter: slowly varying components of $\mu(t)$ are amplified,
while fast fluctuations are attenuated.  

In empirical estimation, assuming $\mu(t)$ to be constant when it is in fact time-varying induces a known bias:
bursts in $\mu(t)$ are misattributed to endogeneity, pushing estimates of $N$ toward criticality \cite{FiliSor15,wheatley2019,wehrli2021}
Allowing for time variation in $\mu(t)$ (e.g., day-by-day calibration or EM/state-space methods) mitigates this
bias and preserves the endo-exo separation.

\subsection{Amplified Memory Parameter}
\label{sec:amplified_tau}

\begin{figure*}[!htb]
	\centering
	\includegraphics[width=0.9\textwidth]{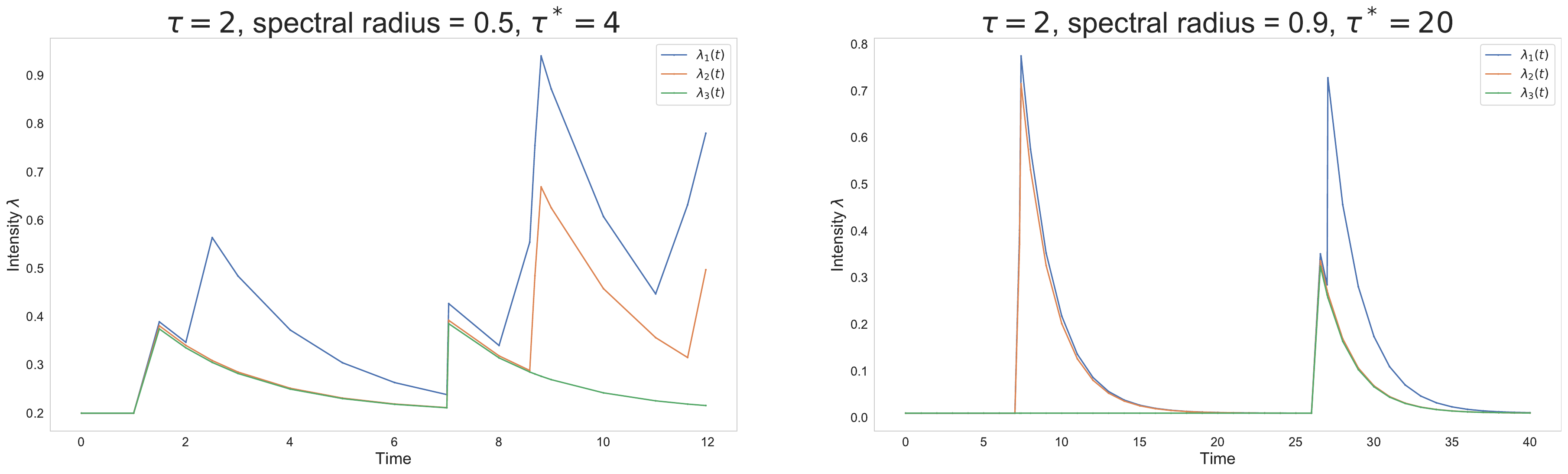}
	\caption{Intensity as a function of  observation time of a simulated Hawkes process with $M=3$ event types,
	for two different values of the spectral radius of the matrix ${N}$ of branching ratios $n_{i,j}$ defined in equation (\ref{dhtyh3wybn2}).
	The left plot shows that a memory time $\tau$ of $2$ is amplified to $4$ for a spectral radius equal to $0.5$.
           The right plot shows that a memory time $\tau$ of $2$ is amplified to $20$ for a spectral radius equal to $0.9$.}
	\label{fig:tau}
\end{figure*}

The kernel $\phi(t)$ introduces memory into the intensity functions $\lambda$ in Equation~\eqref{eq:Hawkes}.  
Building on the previous result that the first moment of the intensity is amplified by the factor $1/(1-n)$ 
(in the case $M=1$, and see expression (\ref{trhjt1ght1}) for $M>1$)  as the spectral radius approaches $1$, 
we now show that memory effects are similarly rescaled.  
Consequently, the characteristic fluctuation time scale of $\lambda$ is of order $\tfrac{\tau}{1-n}$, as illustrated in Figure~\ref{fig:tau}.  
This result can be derived as follows.

Consider a univariate Hawkes process with baseline (exogenous) intensity $\mu(t)$ as a Dirac function $\delta(t)$ corresponding
to a single source event, such that the Laplace transform of $\delta(t)$ is $\delta(\beta) =1$.
The Laplace transform of such a univariate Hawkes process is given by
\begin{equation}
        \tilde{\lambda}(\beta)
        = \tilde{\mu}(\beta) +  n \tilde{\lambda}(\beta) \tilde{\phi}(\beta) 
        = \frac{\tilde{\mu}(\beta)}{1- n\tilde{\phi}(\beta)} 
        = \frac{1}{1- n\tilde{\phi}(\beta)}.
        \label{rh4tb2h2}
\end{equation}
The Laplace transform of an exponential memory kernel $\phi(t) = {\frac{1}{\tau}} e^{-\frac{1}{\tau} t}$ is given by $\tilde{\phi}(\beta) = \frac{1}{\tau\beta + 1}$.
Substituting for $\tilde{\phi}(\beta)$ in (\ref{rh4tb2h2}) gives
\begin{equation}
        \tilde{\lambda}(\beta) = \frac{1}{1- n(\frac{1}{\tau\beta + 1})} = \frac{\tau\beta + 1}{1 -n + \beta \tau}  
         = \frac{1}{1-n} ~\frac{\tau\beta + 1}{1 + \frac{\beta \tau}{1-n}}.
\end{equation}
Since $\frac{\tau\beta + 1}{1 -n + \beta \tau} + \frac{n}{1 -n + \beta \tau} = 1 $,
we can rearrange the above equation as
\begin{equation}
        \tilde{\lambda}(\beta) = 1 + \frac{n}{1-n} \frac{1}{1 + \frac{\tau}{1-n} \beta}.
\end{equation}
The inverse Laplace transform of $\tilde{\lambda}(\beta)$ is then given by:
\begin{equation}
        \lambda(t)
        = \delta(t) + \frac{n}{\tau}e^{- \frac{1-n}{\tau}t} ~.
\end{equation}
The typical relaxation time is thus amplified from the bare time scale $\tau$ of the bare
memory kernel  $\phi(t)$ to the renormalized value $\tau^{\ast} = \frac{\tau}{1-n}$ by the cascade of self-excitations.

Figure~\ref{fig:tau} illustrates the amplification of the effective memory parameter in a three-dimensional Hawkes process simulation.  
In the left panel, the branching ratio matrix $N$ has spectral radius $0.5$ and the baseline memory parameter is $\tau=2$, which yields a renormalized value $\tau^{\ast} = \tfrac{2}{1-0.5} = 4$.  
In the right panel, with spectral radius $0.9$ and the same baseline $\tau=2$, the renormalized memory parameter increases dramatically to $\tau^{\ast} = \tfrac{2}{1-0.9} = 20$, clearly visible as a much slower decay of intensity following a peak of activity.

The amplification of the effective time scale has profound implications for understanding system dynamics.  
As the branching ratio approaches criticality, interactions retain influence over increasingly long horizons, meaning that the relevant time span for observing, modeling, and interpreting the system is itself state-dependent.  
This calls for adaptive evaluation methods that adjust to the effective memory of the system rather than relying on fixed observation windows.  
Ignoring this effect risks truncating long-range dependencies, thereby producing incomplete observations, underestimating causal links, and introducing significant biases in the reconstruction and analysis of networks.

Throughout the manuscript, we express time in units of this renormalized, effective time unit $\tau^{\ast} = \frac{\tau}{1-n}$.

\section{Network Dependency on Construction Parameters}
\label{apx:network_param}

The lead–lag correlation-based emotion network described in the main text depends on two parameters: the bin size $b$, which determines the temporal aggregation window used to construct emotion activity time series, and the lag parameter $\ell$, which specifies how many bins separate the two observations when evaluating lead–lag correlations.
Together, these parameters determine how the underlying event data are discretized and how temporal precedence between emotions is measured.

\begin{figure*}[!htb]
	\centering
	\includegraphics[width=0.9\textwidth]{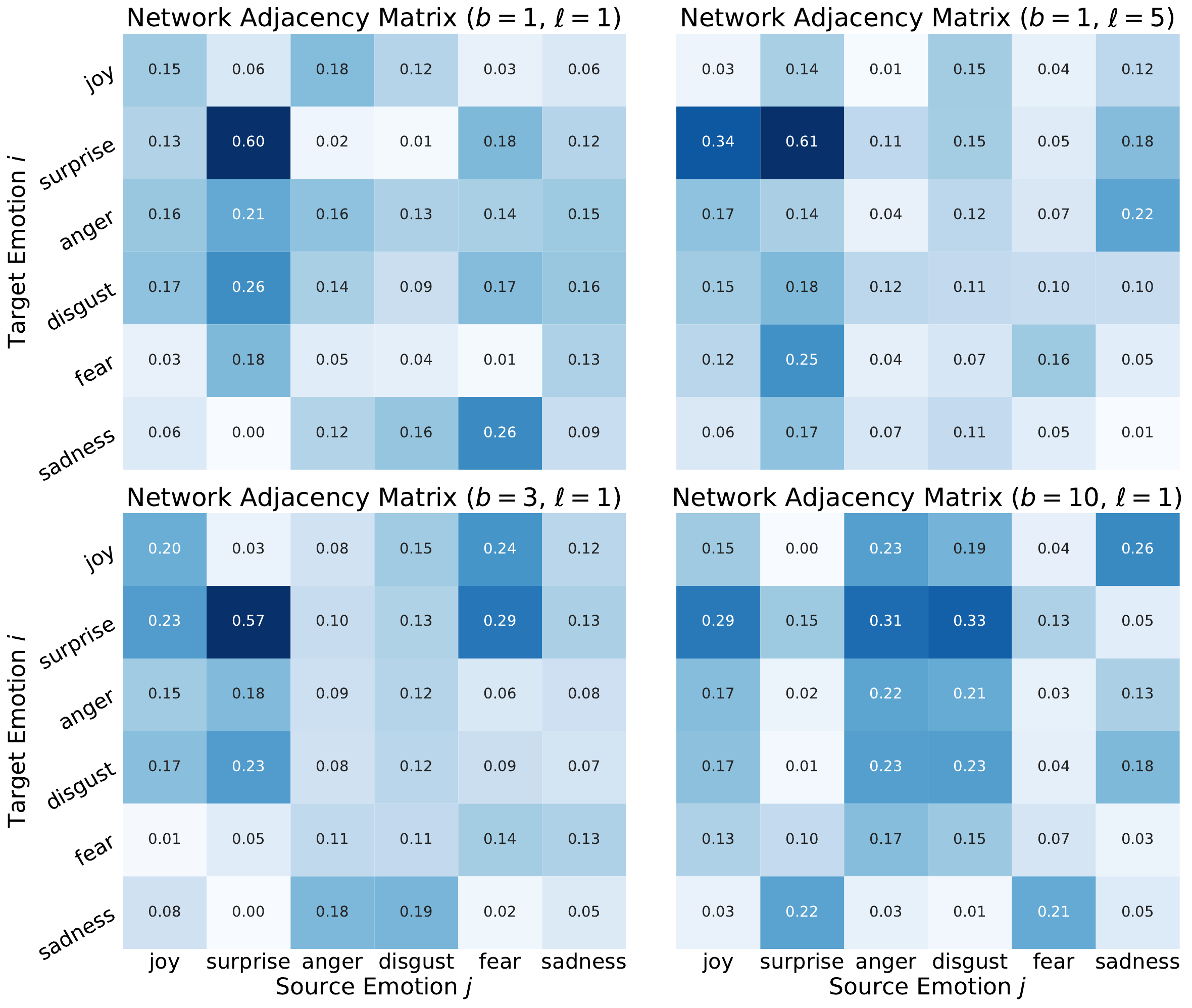}
	\caption{
	Adjacency matrices of correlation-based emotion networks constructed using different combinations of bin size $b$ and lag parameter $\ell$.
	The bin size determines the temporal aggregation window used to construct emotion activity time series, while the lag parameter specifies the number of bins separating the two observations when computing lead–lag correlations.
	Each panel is computed from the same YouTube live-chat data using different parameter values $(b,\ell)$.
	}
	\label{fig:network_param}
\end{figure*}

To illustrate the sensitivity of the resulting networks to these choices, Fig.~\ref{fig:network_param} shows adjacency matrices constructed from the same chat data using different combinations of $b$ and $\ell$.
Although the underlying sequence of messages is identical, the inferred correlations between emotions vary substantially across parameter settings, producing markedly different interaction patterns.
This demonstrates a general limitation of correlation-based network constructions: the inferred adjacency structure can depend strongly on arbitrary modeling choices.

\end{document}